\documentclass[11pt,twoside]{article}
\usepackage{asp2010}
\usepackage{graphicx}

\resetcounters

\begin{document}

\title{Phase-lag Distances of OH Masing AGB Stars}
\author{D.~Engels,$^1$ ~S.~Etoka,$^1$ ~E.~G\'erard$^2$ ~and 
~A.~Richards$^3$
\affil{$^1$Hamburger Sternwarte, Universit\"at Hamburg, Germany}
\affil{$^2$Galaxies, \'Etoiles, Physique et Instrumentation (GEPI), 
Observatoire de Paris, Meudon, France}
\affil{$^3$Jodrell Bank Centre for Astrophysics, Department of Physics 
and Astronomy, University of Manchester, UK} 
}

\begin{abstract}
Distances to AGB stars with optically thick circumstellar shells
cannot be determined using optical parallaxes. However, for stars with
OH 1612 MHz maser emission emanating from their circumstellar shells,
distances can be determined by the phase-lag method. This method
combines a linear diameter obtained from a phase-lag measurement with
an angular diameter obtained from interferometry. The phase-lag of the
variable emission from the back and front sides of the shells has been
determined for 20 OH/IR stars in the galactic disk. These measurements
are based on a monitoring program with the Nan\c{c}ay radio telescope
ongoing for more than 6 years. The inter\-fero\-metric observations are
continuing. We estimate that the uncertainties of the distance
determination will be $\sim$20\%.
\end{abstract}

\vspace*{-6mm}
\section{Introduction}
In the final phases of AGB evolution, the mass-loss rate increase is
such that the star is partially obscured in the visual and even near
infrared bands, as the circumstellar envelope (CSE) gets optically
thick. It is assumed that this occurs mostly for intermediate mass
stars ($M>$\,2\,M$_{\odot}$), but direct evidence is lacking, at least
for the Galactic disk, because of uncertainty in the distances.

Distances to obscured AGB stars cannot be determined using optical
parallaxes; however, the maser emission shown by many OH/IR stars can
be used. Parallax measurements of post-AGB stars and Mira variables
were successfully made using H$_2$O \citep{Imai07} and OH masers
\citep{Vlemmings07}, but have not been attempted for OH/IR stars so
far. Another promising way to obtain reliable distances is the
measurement of the phase-lag between the varying peaks of the OH maser
emission. This yields the absolute linear diameter of the shell and, 
in combination with the angular diameter of the shell measured in
interferometric maps, gives the distance \citep{Herman85a,Langevelde90}.
The OH/IR stars usually exhibit  maser spectra dominated by two peaks
with a separation of $30\pm10$ km\,s$^{-1}$. This double-peak characteristic
can be explained if the maser emission comes from a rather thin,
uniformly expanding shell. The strongest masers then originate from
the front and the back sides of the shell.  The stars are often 
large-amplitude variable stars, and the maser flux density follows these
regular variations. Due to the light travel time through the CSE, the
observer sees the peak from the back side of the shell responding with
a delay $\tau_0$ relative to the peak on the front side. These delays
are of the order of weeks.

With the Nan\c{c}ay Radio Telescope (NRT) we monitored a sample of 20
OH/IR stars to measure their phase-lags. For about half of the sample,
phase-lags had been determined by \citet{Herman85a} and
\citet{Langevelde90}.  Some of the stars served as comparison stars to 
verify the reported phase-lags. For others a re-determination was the 
goal, as the reported phase-lags were inconsistent with each other. 
For the remaining stars, phase-lags were determined for the first time. 
We have also started to image stars at 1612 MHz with the eVLA and 
e-MERLIN near maximum light to determine the angular diameters. We 
report here on the status of the program and give preliminary results.
 
\begin{figure}[h]
\includegraphics[angle=-90,width=6.5cm]{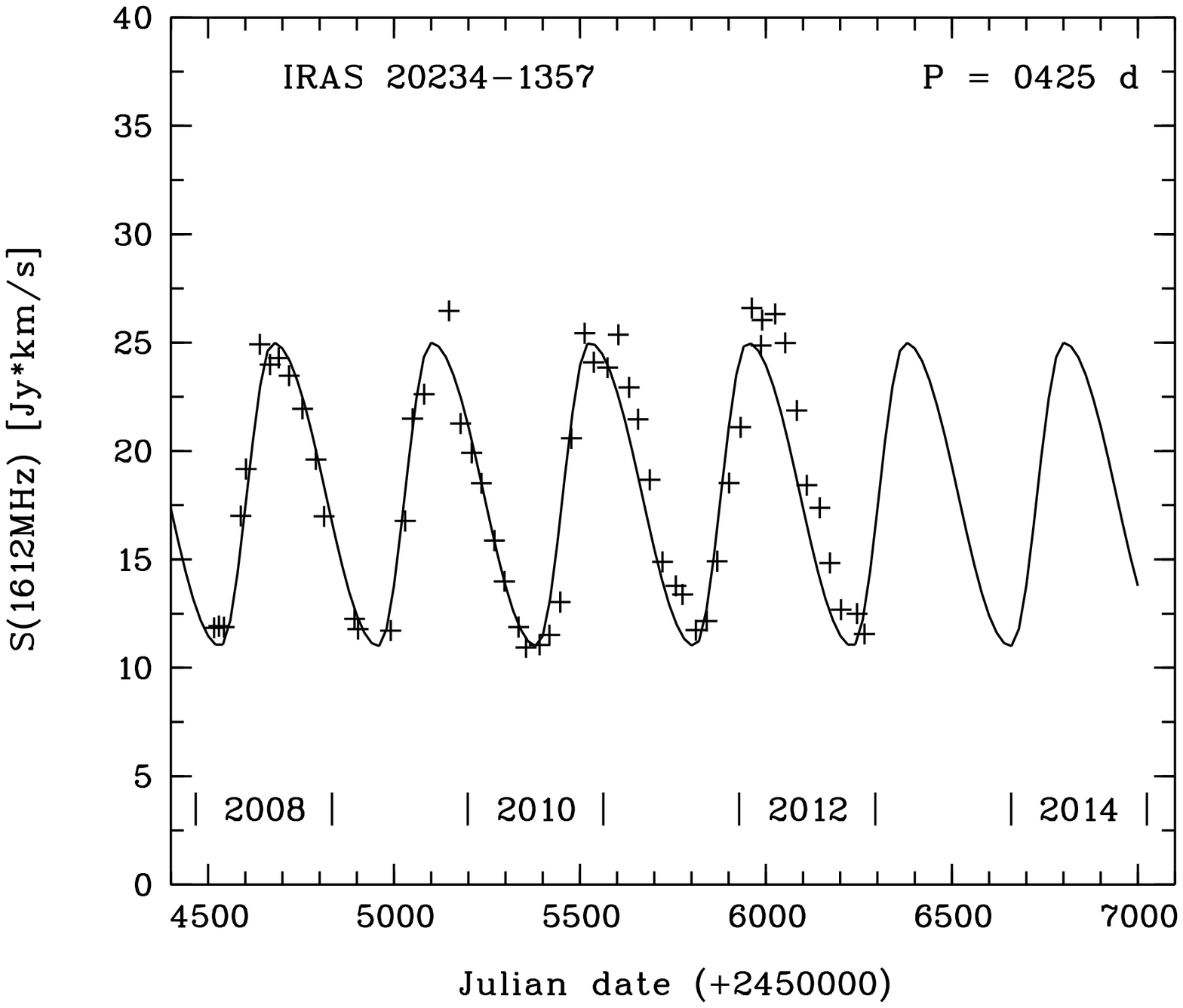} \hspace{1mm}
\includegraphics[angle=-90,width=6.5cm]{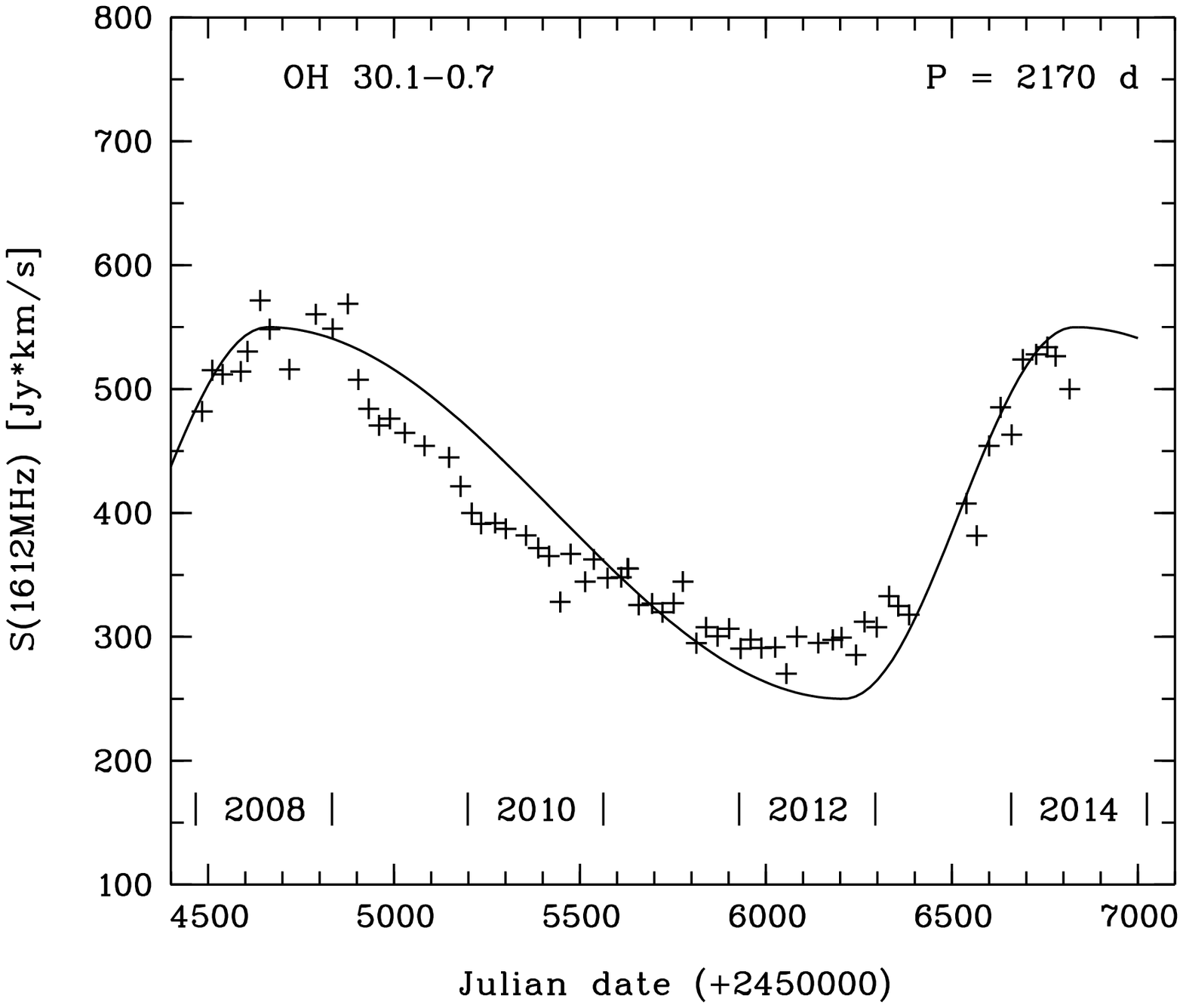}
\caption{Sample light curves of the OH maser flux integrated over the
  full velo\-city profile for IRAS~20234--1357 ($P$ = 1.16 yr) and
  OH~30.1--0.7 ($P$ = 5.95 yr). } 
\label{lightcurves}
\end{figure}

\vspace*{-4mm}
\section{Light Curves and Phase-lag Determinations}

The sample of OH/IR stars was observed with the NRT in the 1612 MHz OH
maser line once every month over 5 years (2008--2012).  For several
sources with insufficient coverage of the variation cycle, the
monitoring is continuing. Typical integration times on source were
5--10 minutes, yielding a typical noise level of 100 mJy. The spectra
were baseline fitted and calibrated (nominal error $\sim5$\%) and
have a velocity resolution of 0.035--0.070~km\,s$^{-1}$.
Representative light curves of the integrated flux are shown in
Figure~\ref{lightcurves}. All stars are large-amplitude variables with
periods $P \approx 1-6$ years. The light curves are smooth with some
month-to-month scatter. This scatter is not due to secular variations
of individual maser features in the spectral profile, superposed on
the smooth variation of the flux integrated over all features. Indeed,
dividing spectra of adjacent months showed that any variations
affected the full profile, and not individual features only. We
therefore conclude that the month-to-month scatter is dominated by
calibration errors.

The light curves of objects with periods $<$\,2 years are well
represented by a sine curve, except that the level of maximum light is
not necessarily recovered in each cycle.  The light curves become
increasingly asymmetric with longer periods, with a steeper rise towards
maximum (duration $0.3-0.4 \cdot P$\,) and a gentle decline towards
minimum. Simple analytical functions, such as an asymmetric sine 
function, are not able to reproduce the detailed shapes of these 
light curves.

\begin{figure}[h]
\includegraphics[angle=-90,width=6.2cm]{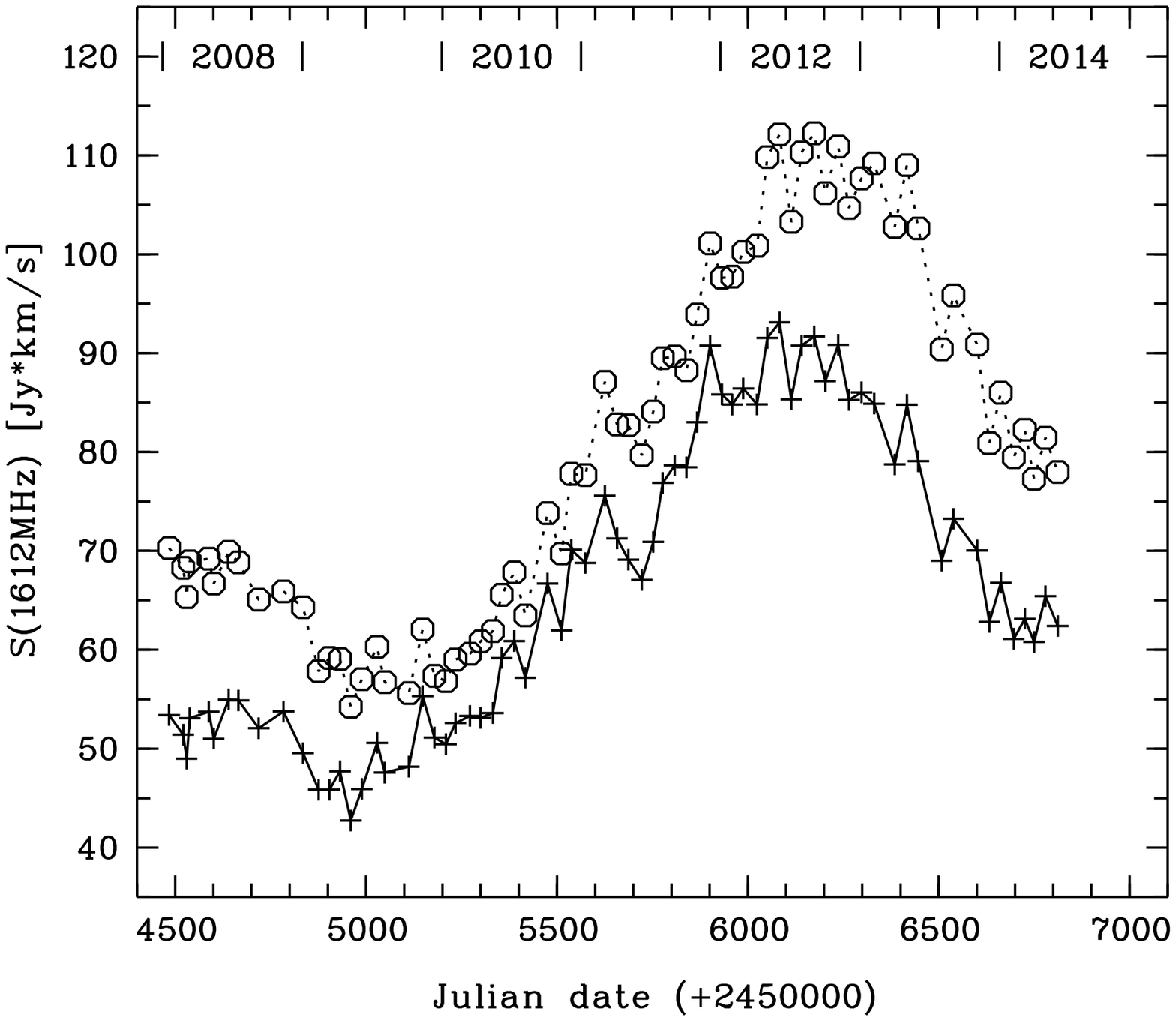} \hspace{6mm}
\includegraphics[angle=-90,width=6.2cm]{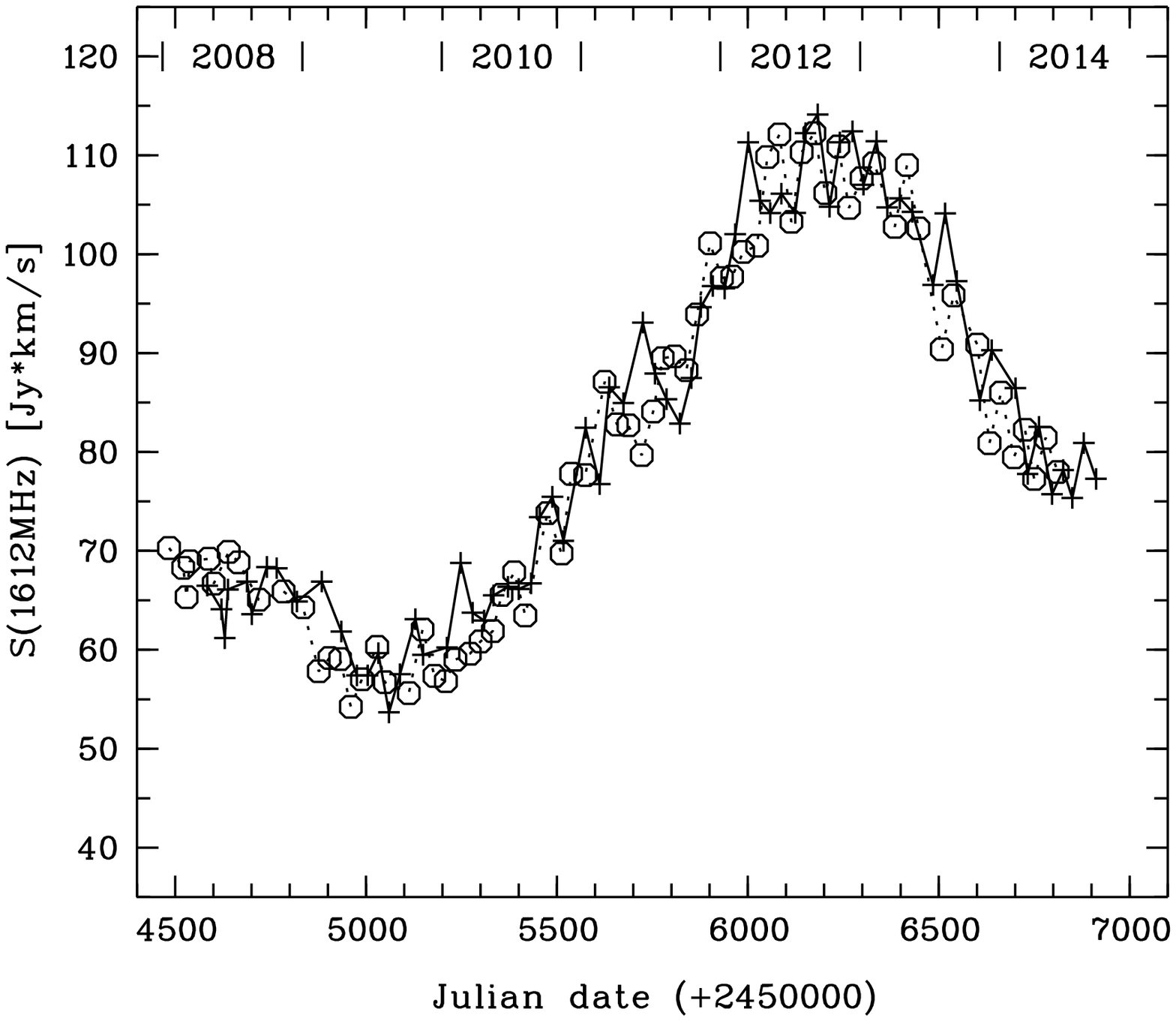}
\caption{{\it Left:} Light curves of the maser emission integrated over
  the blue part ($[-3;+21]$~km~s$^{-1}$; crosses) and the red part
  ($[+21;+45]$~km~s$^{-1}$; circles) of the velocity profile of
  OH~16.1$-$0.3 ($P$ = 6 yrs). {\it Right:} Best match of the
  light curves after scaling and shifting the blue light curve
  and applying a phase-lag $\tau_0 = 100$ days. } 
\label{phaselag}
\end{figure}

To find the phase-lag $\tau_0$ between the front- and back-side
emission (corresponding to the blue and red parts of the maser
velocity profile), we assumed that the maser light curves of both
velocity ranges are similar, e.g.~they differ only in mean flux and
amplitude. For all sources, the peak flux densities of the two maser
peaks were always $\gg$1 Jy, so that all parts of the light curves were
sampled with high signal-to-noise ratio. The phase-lags were then
determined by minimizing the function $\Delta F = F_r(t) - a \cdot
F_b(t+\tau_0) + c$, where $F_b$, $F_r$ are the integrated fluxes over
the blue and red parts of the velocity profile, and $a$ and $c$ are
constants to match amplitude and mean flux of the two maser
features. An example is shown in Figure~\ref{phaselag} for the OH/IR
star OH~16.1$-$0.3, which has one of the longest periods in the sample:
6.03 years. In this case the uncorrected light curves show an apparently
smaller phase-lag in the rising part of the light curve and a larger
lag in the declining part.  This case shows that full coverage of a
variation cycle is essential to determine amplitude and mean flux
correctly before the phase-lag is determined. The resulting phase-lag
of $100\pm10$ days is more than twice as large as the value, 38.3 days,
 measured by \citet{Langevelde90}, although their period of $P=6.2$
years is in good agreement with ours.

\section{Improved Angular Diameters}

The determination of the angular diameter in principle requires
imaging the OH maser shell at tangential velocities, e.g.~in the
center of the velocity interval, where the emission is faint. Because
this emission is hard to detect, the diameters are usually determined
by extrapolation of the projected diameters at non-tangential
velocities assuming that the masers reside in a thin radially symmetric
shell. We are currently observing the stars interferometrically, close
to the maxima of their light curves, to detect as much as possible 
from the faint tangential emission. The aim is to improve the angular
dia\-meters derived from less sensitive observations in the past, to
determine diameters for stars not imaged at 1612 MHz so far, and to
derive constraints on deviations from radial symmetry of the shells. 

OH~16.1$-$0.3 is one of the objects we observed with the eVLA, and
which had not previously been imaged at 1612 MHz. The interferometric
observations were obtained in November 2011 with a velocity resolution
of 0.69~km~s$^{-1}$. Calibration, following the relevant steps for
observations taken in spectral-line mode, and final imaging were
performed with CASA (Common Astronomy Software Application).  A
sensitivity of 2~mJy~beam$^{-1}$ per channel was reached.  The
restoring beam size was ($1\farcs95 \times 1\farcs0$) with a
position angle PA $\approx -28^{\circ}$.  Figure~\ref{fig: VLA
 OH1616.1} presents, in the right panel, the integrated emission 
over the red inner wings of the spectra covering the velocity range 
$[+25;+37]$~km~s$^{-1}$, which gives the best rendering of the diameter 
of the shell, and a typical spectrum of the source at the left. The 
shell is roughly circular with a possible hint of an elongated streak 
of material in the south east of the shell. 

The best fit (by eye) of the projected shell diameter is given by the
magenta circle of $\sim 3\farcs43$ diameter. Correcting for the
projection effect assuming a thin shell, the full angular diameter of
the shell is $\phi = 4\farcs1\pm0\farcs6$.  In combination with the 
linear diameter $d_{\rm{OH}} = (1.73\pm0.17) 10^4$ AU corresponding to
the phase-lag of 100 days, the distance to OH~16.1$-$0.3 is 
$D = 4.2\pm0.7$ kpc.

\begin{figure}[h]
\includegraphics[angle=0,width=13.25cm]{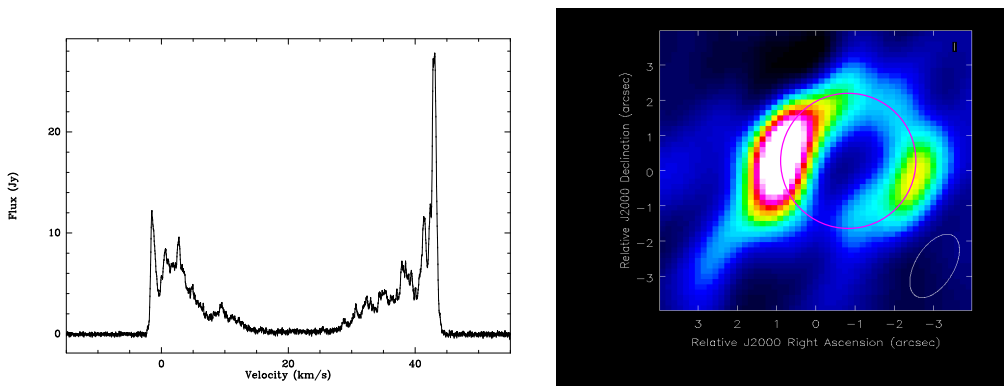} 
\caption{{\it Left:} NRT spectrum of OH~16.1$-$0.3. 
   {\it Right:} Integrated emission over the red inner part of the 
   spectrum of OH~16.1$-$0.3 covering the velocity range 
   $[+27;+37]$~km~s$^{-1}$. The magenta circle presents the best fit 
   of the projected diameter of the shell.} 
\label{fig: VLA OH1616.1}
\end{figure}

\vspace*{-4mm}
\section{Results and Conclusions}

Preliminary \,results \,of \,the \,observing \,program \,of \,the \,full 
\,sample \,are \,summarized \,in Table~\ref{table}.  Given are the 
object name, the newly determined period $P$, the linear dia\-meter 
$d_{\rm{OH}}$ (in AU) $ = 0.1731 \cdot \tau_0$ (in days) calculated 
from the phase-lag $\tau_0$, the angular dia\-meter $\phi$, and the 
distance $D$. Uncertainties are $<3$\% for the periods $P$ 
and approximately 20\% for the linear diameters $d_{\rm{OH}}$. 
$P$ and $d_{\rm{OH}}$ are taken from the NRT observations, while the 
angular diameters of the shell are taken from the literature 
\citep{Amiri12,Baud81,Chapman84,Herman85b,Norris82,Wolak13},
except for OH~16.1$-$0.3 (eVLA) and for OH~83.4$-$0.9 (e-MERLIN),
which come from our own recent observations.

The shell diameters cover a range of $\sim2000 - 17\,000$\,AU, and in
general the dia\-meters are larger for longer periods. The
angular diameters taken from the literature can be rather uncertain,
depending on the degree of extrapolation applied to derive the full
extent of the shells. An obvious case is OH~141.7+3.5, where the
angular diameter provided by \citet{Chapman84} is strongly
underestimated, yielding an unreasonably great distance. The accuracy
of the distances currently is not better than 20\%, where this 
uncertainty reflects only the uncertainty of the phase-lag determination. 

\begin{table} 
{\scriptsize
\caption[ ]{Results from the OH 1612 MHz monitoring program.  
\label{table}}
\begin{flushleft}
\begin{tabular}{lrrrr|lrrrr}
\noalign{\smallskip}\hline\noalign{\smallskip}
  Object  & $P$~~ & $d_{\rm{OH}}$~~~  &  $\phi$~ &  $D$~~~  &  Object  & 
$P$~~ & $d_{\rm{OH}}$~~~   &  $\phi$~~ &  $D$~~~ \\
    & [yr]  & [$10^3$ AU] & [$^{\prime\prime}$]  & [kpc] &  & [yr] & [$10^3$ AU] &  [$^{\prime\prime}$]~  & [kpc]  \\
\noalign{\smallskip}\hline\noalign{\smallskip}
20234$-$1357   & 1.16  &$<1.7$~~~ &   &    &  OH 32.0$-$0.5 & 4.16  & 11.2~~~ & 1.07 & 10.6~  \\             
OH 44.8$-$2.3  & 1.47  & 2.6~~~  & 2.6 & 1.0~  &  OH 127.8+0.0  & 4.36  & 9.5~~~  & 2.69 & 3.6~   \\  
IRC +50137     & 1.74  & 1.7~~~  &    &    &  OH 26.5+0.6   & 4.36  & 6.0~~~  & 4.49 & 1.4~  \\   
01037+1219     & 1.78  & 4.3~~~  & 8.0 & 0.5~  &  OH 75.3$-$1.8 & 4.52  & 10.4~~~ &      &       \\  
05131+4530     & 2.88  & 5.2~~~  &    &    &  OH 32.8$-$0.3 & 4.63  & 10.4~~~ & 2.74 & 3.8~  \\
OH 39.7+1.5    & 3.45  & 2.9~~~  & 4.0 & 0.7~  &  OH 20.7+0.1   & 4.71  & 15.6~~~ & 1.65 & 9.4~  \\  
OH 55.0+0.7    & 3.48  & 8.6~~~  &    &    &  OH 104.9+2.4  & 4.79  & 6.0~~~  & 2.88 & 2.1~  \\
21554+6204     & 3.51  & 5.2~~~  &    &    &  OH 30.1$-$0.7 & 5.95  & 10.4~~~ & 2.69 & 3.9~  \\  
OH 138.0+7.2   & 3.86  & 1.7~~~  & 0.8 & 2.2~  &  OH 16.1$-$0.3 & 6.03  & 17.3~~~ & 4.10  & 4.2~  \\
OH 83.4$-$0.9  & 4.11  & 5.2~~~  & $\sim$1.8 & $\sim$3.0~ &  OH 141.7+3.5  & 6.05  & 13.0~~~ & 0.76 & 17.1:  \\ 
\noalign{\smallskip}\hline
\end{tabular}
\end{flushleft}
}
\end{table}

For eleven stars of the sample, phase-lags (linear shell diameters) are
available from \citet{Langevelde90}. Within our estimated
uncertainty of 20\% we obtained agreement for OH~26.5+0.6,
OH~32.0$-$0.5, OH~44.8$-$2.3 and OH~104.9+2.4. For the remaining
sources, as in the case of OH~16.1$-$0.3, higher deviations  were
found. OH~127.8+0.0 was monitored by \citet{Wolak13} between 2002
and 2008, also with the NRT. They determined a period of 1600 days and
a phase-lag of 62 days, which is in perfect agreement with our results
($P=1590$ days and $\tau_0 = 55$ days).

We expect that the linear shell diameters can be determined with an
error $\sim$10\%, by analyzing the variations of the phase-lag
across the velocity profile \citep[see][]{Wolak13}, instead of using
the integrated fluxes. In addition, the uncertainties in the distances
will depend on the deviations from radial symmetry of the shells. 
Assuming that these deviations do not exceed 10\% \citep{Etoka10}, 
the uncertainties in the distances should not exceed 20\%. This will 
be probed by the interferometric observations, which are expected to 
continue until 2017.
 
\bibliography{b_engels}

\question{Whitelock} Very nice results. Do you have accurate luminosities 
from the mid-infrared to go with the distances?

\answer{Engels} Bolometric fluxes can be obtained nowadays between 2 and 
200 $\mu$m by various IR surveys. Combining them with the phase-lag 
distances gives a range of 5000 -- 50\,000\,L$_{\odot}$, consistent with 
a classification as intermediate-mass stars at the very end of AGB 
evolution.

\question{Olofsson} If your stars are in the super-wind phase, it appears 
to me that you find much larger durations of this phase than the ones 
obtained from dust studies (see Lombaert's presentation).

\answer{Engels} The masers are pumped by photons emitted in the superwind 
at much smaller distances from the stars than the masers themselves. 
The masers might be located at distances where the dust and gas is from 
the pre-superwind phase.

\end{document}